\documentclass[final]{article}
\begin{document}

\title{\textbf{Quasiprobability and Probability Distributions for Spin $1/2$ States}\footnote{Accepted by \textrm{Foun. Phys. Lett.}, copyright\copyright by Kluwer Academic / Plenum Publishers.}}
\author{M.O. Terra Cunha$^{\dag}$, V.I. Man'ko$^{\ddag}$, and M.O. Scully$^{\S}$ \and 
\and $^{\dag}$ Instituto de Ci\^{e}ncias Exatas \\
Universidade Federal de Minas Gerais \\
CP 702, 30123-970, Belo Horizonte, MG, Brazil \\
\textsf{tcunha@mat.ufmg.br}
\and $^{\ddag}$ P. N. Lebedev
Physical Institute \\
Russian Academy of Sciences \\
Leninskii Pr. 53, Moscow 117924, Russia 
\and $^{\S}$ Max Plank Institut
f\"{u}r Quantenoptik \\
85748 Garching, Germany \\
and \\
Department of Physics and Institute for Quantum Studies \\
Texas A \& M University College Station, Texas 77843, USA}

\maketitle

\begin{abstract}
\ We develop a Radon like transformation, in which $P$ quasiprobability distribution for spin $1/2$ states is written in terms of the tomographic probability distribution $w$.
\end{abstract}

\section*{Introduction}

The idea of finding a description of quantum states in terms of
(generalized) functions which are analogous to probability distributions in
particle phase space was realized by Wigner\cite{Wig32} 
(see also ref.\cite{HOSW84}). The Wigner function, which is a real valued distribution, turned
out to take negative values in some domains of phase space for some quantum
states. Due to this the Wigner function was called a quasiprobability
distribution (or a quasidistribution). The quantum mechanical basic
equations were rewritten in terms of the Wigner function by Moyal\cite{Moy49}. Feynman\cite{Fey87} has even discussed the possibility to drop the
assumption that the probabilities of an event must be a nonnegative number.
These attempts are compared with the permanent wish to explain quantum
mechanics in intuitively more acceptable classical notions. The hidden
variables theories are examples of attempts in this direction\cite{Bel73}.

Addressing the question of how to make quantum mechanics look like a hidden
variable theory and vice versa, an analog of Glauber-Sudarshan\cite{Gla63-Sud63} $P$ distribution was constructed for spin $1/2$ states by one
of the authors\cite{Scu83}. The introduced quasiprobability distribution
was expressed in terms of the conventional density matrix description of
spin $1/2$ states. This distribution was applied to propose a hidden
variable theory, whose predictions agree with quantum mechanical ones in
various cases (for a tutorial, see ref.\cite{ScuMex}). This technique was generalized to any pair of noncommuting
observables and used to reinterpret Einstein-Podolsky-Rosen-Bohm ``paradox''\cite{CS86}, and the role of Feynman's negative probabilities was made
concrete in the example of interferometers with which path (WW for the
German \textit{welcher Weg}) detectors\cite{SWS94}. 
Furthermore, the question of operator ordering was addressed in ref.\cite{Cha92}, where some analogous distributions were proposed.

Recently the tomographic methods of measuring quantum states were dis\-cussed\cite{Ber87-VR89-SBRF93-MMT96-7}. In these schemes the invertible Radon transform was used to express the Wigner function in terms of positive probability distributions (called marginal distributions). Such marginal
distributions can be obtained directly from measurements, and are used to completely characterize a quantum state. The tomography of spin states was discussed in refs.\cite{DM97-MM97-AMMS98}. On the other hand, for spin $1/2$ states the relation between quasiprobability $P$ distribution\cite{Scu83} and positive probability distribution\cite{DM97-MM97-AMMS98} has not been discussed yet.

The aim of our work is to find the map which is analog of Radon transform used in tomographic schemes\cite{Ber87-VR89-SBRF93-MMT96-7} for connecting the quasiprobability distribution for spin $1/2$ particles with measurable 
probability distributions. We also express the density matrix in terms of the distributions $P$ and $w$, and relate the last with Bloch vector.

This Letter is organized as follows: in section \ref{RevP} the properties of $P$ quasiprobability distribution are reviewed; in section \ref{Spinprob} the $w$ probability distribution is discussed. In section \ref{Pw} their relation is shown. The paper ends with a few concluding remarks and some calculational tools are collected in the appendix.

\section{A Review on $P\left( \vec{s}\right) $ Spin $1/2$
Quasiprobability Distribution}

\label{RevP}Some years ago, one of the authors has introduced a quasiprobability distribution $P\left( \vec{s}\right) $ in order to describe a spin $1/2$ quantum state\cite{Scu83}. In some features it can 
be viewed as an analog of Glauber-Sudarshan $P$ distributions for light (equivalently harmonic oscillator) quantum states\cite{Gla63-Sud63}. For a tutorial, the reader is referred to ref.\\cite{ScuMex}.

In order to define $P\left( \vec{s}\right) $ we should first introduce the $\delta $ operator given by 
\[
\delta \left( c-\widehat{\mathcal{O}}\right) \equiv \smallint \frac{d\chi }{2\pi }\exp \left[ -i\chi \left( c-\widehat{\mathcal{O}}\right) \right] , 
\]
where $\widehat{\mathcal{O}}$ is an operator (e.g. Pauli spin operator $\widehat{\sigma _z}$) and $c$ is a real variable which plays the role of its classical analog (e.g. $s_z$). Then, given an operator $\widehat{Q}\left( 
\vec{\sigma }\right) $, we can define a classical counterpart $Q\left( \vec{s}\right) $ by 
\[
\widehat{Q}\left( \vec{\sigma }\right) =\smallint d^3sQ\left(
s_x,s_y,s_z\right) \delta \left( s_x-\widehat{\sigma _x}\right) \delta
\left( s_y-\widehat{\sigma _y}\right) \delta \left( s_z-\widehat{\sigma _z}
\right) , 
\]
where should be noticed that an operator ordering was chosen. Quantum mean value 
\begin{equation}
\left\langle \widehat{Q}\right\rangle =Tr\left( \rho \widehat{Q}\right),
\label{qmv} 
\end{equation}
and ``classical'' mean value 
\begin{equation}
\left\langle \widehat{Q}\right\rangle =\smallint d^3sP\left( \vec{s}\right) Q\left( \vec{s}\right), 
\label{cmv}
\end{equation}
suggest the definition
\begin{equation}
P\left( \vec{s}\right) =Tr\left[ \rho \delta \left( s_x-\widehat{\sigma _x}\right) \delta \left( s_y-\widehat{\sigma _y}\right) \delta \left(s_z-\widehat{\sigma _z}\right) \right] , 
\label{P}
\end{equation}
which makes quantum and ``classical'' mean values (eqs.\ (\ref{qmv}) and (\ref{cmv})) to coincide.

For the purposes of ref.\cite{Scu83} (treat Stern-Gerlach like experiments) the referred author has ``traced out'' the $y$ variable. We will not proceed in this way! We will treat $P\left( \vec{s}\right) $ in its full 
glory, in the same spirit as in ref.\cite{Cha92}, where some important properties are made more evident.

\subsection{An Equivalent $P\left( \vec{s}\right) $ Presentation}

With a few algebraic manipulation, we get another expression for $P\left( \vec{s}\right) $, in which some important features easily appears.

Just inserting the definitions of $\delta \left( m_i-\widehat{\sigma _i}\right) $ and using eigenvectors of $\widehat{\sigma _z}$ to calculate trace, we get 
\begin{equation}
P\left( \vec{s}\right) =\sum_{a}\smallint \frac{d\chi }{2\pi }\smallint \frac{d\zeta }{2\pi }\smallint \frac{d\eta }{2\pi }
e^{-i\chi s_x}e^{-i\zeta s_y}e^{-i\eta s_z}\left\langle a_z\right| \rho e^{i\chi \widehat{\sigma _x}}e^{i\zeta \widehat{\sigma _y}}e^{i\eta \widehat{\sigma _z}}\left| a_z\right\rangle ,  \label{Pdistr}
\end{equation}
where 
\[
\widehat{\sigma _z}\left| a_z\right\rangle =a\left| a_z\right\rangle . 
\]

Standard quantum mechanical calculation immediately leads to 
\begin{eqnarray}
\left\langle a_z\right| \rho e^{i\chi \widehat{\sigma _x}}e^{i\zeta \widehat{\sigma _y}}e^{i\eta \widehat{\sigma _z}}\left| a_z\right\rangle &=&\sum_{b,c}e^{i\eta a}e^{i\zeta b}e^{i\chi c}
\label{matelem} \\
&&\left\langle c_x\mid b_y\right\rangle \left\langle b_y\mid
a_z\right\rangle \left\langle a_z\left| \rho \right| c_x\right\rangle , 
\nonumber
\end{eqnarray}
where 
\begin{eqnarray}
\widehat{\sigma _x}\left| c_x\right\rangle &=&c\left| c_x\right\rangle , \nonumber \\
\widehat{\sigma _y}\left| b_y\right\rangle &=&b\left| b_y\right\rangle . \nonumber
\end{eqnarray}

Substitution of eq.\ (\ref{matelem}) in eq.\ (\ref{Pdistr}) yields 
\[
P\left( \vec{s}\right) =\sum_{a,b,c} \delta ^3\left( \vec{s}-\left( c,b,a\right) \right) \left\langle c_x\mid
b_y\right\rangle \left\langle b_y\mid a_z\right\rangle \left\langle a_z\left| \rho \right| c_x\right\rangle , 
\]
where we made use of the notation 
\[
\delta ^3\left( \vec{s}-\left( c,b,a\right) \right) \equiv \delta \left( s_x-c\right) \delta \left( s_y-b\right) \delta \left( s_z-a\right) . 
\]
Decomposition of $\left| c_x\right\rangle $ in $\left| a_z\right\rangle $ basis leads to
\begin{eqnarray}
P\left( \vec{s}\right) &=&\sum_{a,b,c}\delta ^3\left( \vec{s}-\left( c,b,a\right) \right) \left\langle c_x\mid
b_y\right\rangle \left\langle b_y\mid a_z\right\rangle  \label{Pinvertable}
\\
&&\ \left\{ \left\langle a_z\mid c_x\right\rangle \left\langle a_z\left| \rho \right| a_z\right\rangle +\left\langle -a_z\mid c_x\right\rangle
\left\langle a_z\left| \rho \right| -a_z\right\rangle \right\} .  \nonumber
\end{eqnarray}

The above expression (which corresponds to eq.(2.18) in ref.\cite{Cha92}, but written in coordinates) makes clear the singular character of $P\left( \vec{s}\right) $ for quantum mechanical allowed states. It can assume non null values only in the eight vertices of a cube centered at the origin and parallel to the Cartesian axis. This is an obvious manifestation of the fact that only eigenvalues are allowed as results of a single measurement in quantum mechanics.

We make use of this singular character and define $p\left( c,b,a\right) $ by 
\begin{equation}
P\left( \vec{s}\right) =\sum_{a,b,c}\delta ^3\left( \vec{s}-\left( c,b,a\right) \right) p\left( c,b,a\right) .
\label{p} 
\end{equation}
Straightforward calculation (see appendix) leads to 
\begin{eqnarray}
p\left( 1,1,1\right) &=&\frac 14\left( 1+i\right) \left[ \rho _{++}+\rho _{+-}\right] ,  \label{pexplict} \\
p\left( -1,1,1\right) &=&\frac 14\left( 1-i\right) \left[ \rho _{++}-\rho _{+-}\right] ,  \nonumber \\
p\left( 1,-1,1\right) &=&\frac 14\left( 1-i\right) \left[ \rho _{++}+\rho _{+-}\right] ,  \nonumber \\
p\left( -1,-1,1\right) &=&\frac 14\left( 1+i\right) \left[ \rho _{++}-\rho _{+-}\right] ,  \nonumber \\
p\left( 1,1,-1\right) &=&\frac 14\left( 1-i\right) \left[ \rho _{--}+\rho _{-+}\right] ,  \nonumber \\
p\left( -1,1,-1\right) &=&\frac 14\left( 1+i\right) \left[ \rho _{--}-\rho _{-+}\right] ,  \nonumber \\
p\left( 1,-1,-1\right) &=&\frac 14\left( 1+i\right) \left[ \rho _{--}+\rho _{-+}\right] ,  \nonumber \\
p\left( -1,-1,-1\right) &=&\frac 14\left( 1-i\right) \left[ \rho _{--}-\rho _{-+}\right] ,  \nonumber
\end{eqnarray}
where we used the simple notation 
\begin{eqnarray}
\rho _{++} &=&\left\langle +_z\left| \rho \right| +_z\right\rangle ,\nonumber \\
\rho _{+-} &=&\left\langle +_z\left| \rho \right| -_z\right\rangle ,\nonumber \\
\rho _{-+} &=&\left\langle -_z\left| \rho \right| +_z\right\rangle ,\nonumber \\
\rho _{--} &=&\left\langle -_z\left| \rho \right| -_z\right\rangle .\nonumber
\end{eqnarray}

\subsection{Density Matrix from $P\left( \vec{s}\right) $}

We now show how to invert relation (\ref{Pinvertable}) and obtain density matrix from $P\left( \vec{s}\right) $ quasiprobability distribution. Then, with aid of simple properties of density matrices, we
derive some properties which a $P\left( \vec{s}\right) $ distribution should obey to correspond to a quantum mechanical possible state.

From the first two lines of eqs.\ (\ref{pexplict}) one easily obtains 
\begin{eqnarray}
\rho _{++} &=&\left( 1-i\right) p\left( 1,1,1\right) +\left( 1+i\right) p\left( -1,1,1\right) ,\label{rhofromp} \\
\rho _{+-} &=&\left( 1-i\right) p\left( 1,1,1\right) -\left( 1+i\right) p\left( -1,1,1\right) ,\nonumber
\end{eqnarray}
and using the properties of hermiticity and unitary trace of density matrix, one gets 
\[
\rho =\left[ 
\begin{array}{cc}
\rho _{++} & \rho _{+-} \\ 
\rho _{+-}^{*} & 1-\rho _{++}
\end{array}
\right] . 
\]

This result shows us that from an information theoretical viewpoint all other six $p\left( c,b,a\right) $ are redundant information. Algebraically, this redundant information manifests itself as a set of constraints which a $ P\left( \vec{s}\right) $ distribution must obey, in order to represent a quantum mechanical allowed state (\textit{e.g.} one which obeys uncertainty relations). One of this constraints is just marginal distribution property: 
\[
\sum_{b,c}p\left( c,b,a\right) =\left\langle a_z\left| \rho \right| a_z\right\rangle =\mathcal{P}\left( s_z=a\right) , 
\]
with similar relations holding for $\mathcal{P}\left( s_x=c\right) $ and $ \mathcal{P}\left( s_y=b\right) $, where $\mathcal{P}\left( *\right) $ denotes probability of $*$. So now it is clear that $P\left( \vec{s}\right) $ is a complex valued quasiprobability distribution.

\subsection{Examples}

We now work out some simple examples.

\textit{i}) Spin up in $z$-axis:

As its density matrix in $\left| a_z\right\rangle $ basis is given by 
\[
\rho _{+z}=\left[ 
\begin{array}{cc}
1 & 0 \\ 
0 & 0
\end{array}
\right] , 
\]
one immediately obtains 
\begin{eqnarray}
p\left( 1,1,1\right) &=&p\left( -1,-1,1\right) =\frac 14\left( 1+i\right) ,\nonumber \\
p\left( -1,1,1\right) &=&p\left( 1,-1,1\right) =\frac 14\left( 1-i\right) ,\nonumber \\
p\left( c,b,-1\right) &=&0.\nonumber
\end{eqnarray}

This example was also worked out in ref.\cite{Scu83}, but considering only $x$ and $z$ directions. If we add all $y$ possibilities ($b=\pm 1$) we reobtain that result.

\textit{ii}) Spin ``up'' in $x$-axis:

Equivalently, the density matrix is given by 
\[
\rho _{+x}=\left[ 
\begin{array}{cc}
1/2 & 1/2 \\ 
1/2 & 1/2
\end{array}
\right] , 
\]
and then 
\begin{eqnarray}
p\left( 1,1,1\right) &=&p\left( 1,-1,-1\right) =\frac 14\left( 1+i\right) ,\nonumber \\
p\left( 1,-1,1\right) &=&p\left( 1,1,-1\right) =\frac 14\left( 1-i\right) ,\nonumber \\
p\left( -1,b,a\right) &=&0.\nonumber
\end{eqnarray}

\textit{iii}) Spin ``up'' in $y$-axis:

Density matrix is given by 
\[
\rho _{+y}=\left[ 
\begin{array}{cc}
1/2 & -i/2 \\ 
i/2 & 1/2
\end{array}
\right] , 
\]
and then 
\begin{eqnarray}
p\left( c,1,a\right) &=&\frac 14\nonumber \\
p\left( 1,-1,1\right) &=&p\left( -1,-1,-1\right) =\frac {-i}{4},\nonumber \\
p\left( -1,-1,1\right) &=&p\left( 1,-1,-1\right) =\frac i4\nonumber .
\end{eqnarray}

Those examples show how peculiar is the behaviour of $P\left( \vec{s}\right) $
quasiprobability with respect to cube symmetries. This is a consequence of the noncommutativity of $\delta $ operators in eq.\ (\ref{P}), which makes $P\left( \vec{s}\right) $ distribution noncovariant.
In case we had used the symmetric Margenau-Hill $P$, $P_s\left( \vec{s} \right) $, or the Wigner-Weyl $P$, $P_w\left( \vec{s} \right) $, as defined in ref.\cite{Cha92}, they would have a covariant behaviour, but in this work we intend to discuss the original $P\left( \vec{s} \right) $ distribution (denoted $P_{xyz}$ in ref.\cite{Cha92}). 

\textit{iv}) Unpolarized spin:

In this last example, density matrix is given by 
\[
\rho _o=\left[ 
\begin{array}{cc}
1/2 & 0 \\ 
0 & 1/2
\end{array}
\right] , 
\]
and one gets 
\[
\begin{array}{ccccc}
p\left( 1,1,1\right) &=&p\left( -1,-1,1\right) &=&\\
p\left( -1,1,-1\right) &=&p\left( 1,-1,-1\right) &=&\frac 18\left( 1+i\right), \\
p\left( -1,1,1\right) &=&p\left( 1,-1,1\right) &=&\\
p\left( 1,1,-1\right) &=&p\left( -1,-1,-1\right) &=&\frac 18\left( 1-i\right).
\end{array}
\]

Those four examples span all possibilities, since any spin $1/2$ density operator can be written as
\begin{equation}
\rho = \frac 12\left\{ {\widehat{1}} + \left\langle \widehat{\sigma _x}\right\rangle \widehat{\sigma _x} + \left\langle \widehat{\sigma _y}\right\rangle \widehat{\sigma _y} + \left\langle \widehat{\sigma _z}\right\rangle \widehat{\sigma _z} \right\},
\label{rhofrommv}
\end{equation}
and the relation between $P\left( \vec{s}\right) $ and $\rho $ is linear.

\section{Spin States in the Probability Representation}

\label{Spinprob}We review in this section the probability representation of spin $1/2$ states. This representation was introduced in refs.\cite{DM97-MM97-AMMS98} and treats both, pure states and statistical mixtures equally via probability distributions. For a general review, see ref.\cite{ManMex}.

The central point in such a representation is the property of density matrices that, in whatever basis they are written, theirs diagonal elements are nonnegative probabilities. If a spin state is given by a density matrix $\rho ^{\left( 1/2\right) }$ (written with respect to the $\widehat{\sigma _z}$ eigenvectors), one can describe the same state by a rotated reference frame density matrix $\rho ^{\left( 1/2\right) }\left( u\right) $, where $u$ is the set of Euler angles $\phi $, $\theta $, $\psi $, which defines the rotation (as usual $0\leq \phi ,\psi <2\pi $, $0\leq \theta \leq \pi $).

Let us describe in some detail how such a rotation is done. For the sake of simplicity, let us work first with a pure state, described by a spinor $\Psi =\left( 
\begin{array}{c}
\chi _{+} \\ 
\chi _{-}
\end{array}
\right) $, again with respect to the $\widehat{\sigma _z}$ eigenvectors. In a rotated reference frame, the same state is described by the rotated spinor given by
\begin{equation}
\Psi \left( u\right) =\mathbf{D}\left( u\right) \Psi ,  \label{rotspin}
\end{equation}
where
\[
\mathbf{D}\left( u\right) =\left[ 
\begin{array}{cc}
\cos \frac \theta 2e^{i\left( \phi +\psi \right) /2} & \sin \frac \theta
2e^{-i\left( \phi -\psi \right) /2} \\ 
-\sin \frac \theta 2e^{i\left( \phi -\psi \right) /2} & \cos \frac \theta
2e^{-i\left( \phi +\psi \right) /2}
\end{array}
\right] 
\]
is an irreducible spin $1/2$ matrix representation of the rotation group\cite{Ham}. In components eq.\ (\ref{rotspin}) reads
\[
\chi _s\left( u\right) =\sum_mD_{sm}^{\left( 1/2\right) }\left( u\right) \chi _m,
\]
and it is worthy to note that the matrix elements $D_{sm}^{\left( 1/2\right) }\left( u\right) $ are also named Wigner functions (for rotations)\cite{Sak}.

It is now straightforward to pass to density matrix. The same state is described by 
\[
\rho ^{\left( 1/2\right) }=\Psi \Psi ^{\dagger },
\]
and rotation gives
\begin{eqnarray}
\rho ^{\left( 1/2\right) }\left( u\right)  &=&\Psi \left( u\right) \left( \Psi \left( u\right) \right) ^{\dagger }=\mathbf{D}\left( u\right) \Psi \left( \mathbf{D}\left( u\right) \Psi \right) ^{\dagger }  \label{rotrho} \\
&=&\mathbf{D}\left( u\right) \Psi \Psi ^{\dagger }\left( \mathbf{D}\left( u\right) \right) ^{\dagger }=\mathbf{D}\left( u\right) \rho ^{\left( 1/2\right) }\left( \mathbf{D}\left( u\right) \right) ^{\dagger }.  \nonumber
\end{eqnarray}
As eq.\ (\ref{rotrho}) is linear in $\rho ^{\left( 1/2\right) }$, it also applies for mixed states.

The nonnegative matrix elements $\rho _{ii}^{\left( 1/2\right) }\left( u\right) $ have the meaning of probability distribution of obtaining $i$ when the rotated $z$-axis spin component is measured in this state. By the
geometry of Euler angles it is clear that $\rho _{ii}^{\left( 1/2\right) }\left( u\right) $ do not depend on $\psi $, so $u\simeq \left( \theta ,\phi \right) $ can be viewed as points in a sphere (equivalently, $\rho _{ii}^{\left( 1/2\right) }\left( u\right) $ does not depend on the rotation itself, but only on the new oriented quantization axis, defined by $\left( \theta ,\phi \right) $). The central object in this approach is then defined 
\begin{equation}
w\left( i,u\right) \equiv \rho _{ii}^{\left( 1/2\right) }\left( u\right) , \label{wfromro}
\end{equation}
which is a positive distribution obeying 
\begin{equation}
\sum_{i}w\left( i,u\right) =1,
\label{marginal}
\end{equation}
for all $u$ (i.e.: marginal probabilities property). It should be viewed as a probability distribution for spin $1/2$ states.

\subsection{Density Matrix from $w\left( i,u\right) $}

In refs.\cite{DM97-MM97-AMMS98} $w\left( i,u\right) $ probability distributions are worked out for arbitrary spin states. With aid of some general properties of $SU\left( 2\right) $ group, inverse integral transformations are obtained, and density matrix is given in terms of $w\left( i,u\right) $. As in the present work we are only interested in spin $1/2$ states, we will use a more direct tomographic approach.

As spin Hilbert Spaces are finite dimensional, the complete description of physical state can be achieved by using a finite set of real numbers. In particular, for two level systems, three real numbers are necessary and sufficient to completely describe an arbitrary state. Tomographically, it means that three well chosen axis can give the complete knowledge of the state of the system. A manifestation of these facts, with a canonical choice of axis, is given by  eq.\ (\ref{rhofrommv}).

The mean values involved in eq.\ (\ref{rhofrommv}) can be directly obtained by the definition of $w\left( i,u\right) $ distribution, since
\begin{equation}
\left\langle \widehat{\sigma _u} \right\rangle =
w\left( +\frac12,u\right) - w\left( -\frac12,u\right) ,
\nonumber
\end{equation}
where $\widehat{\sigma _u} \equiv \vec {\widehat{\sigma }} \cdot \vec{u}$, and marginal probability property (eq.\ (\ref{marginal})). One obtains
\begin{equation}
\left\langle \widehat{\sigma _u} \right\rangle = 2 w\left( +\frac12,u\right) - 1.
\label{mvfromw+}
\end{equation}
Substitution of eqs.\ (\ref{mvfromw+}) in eq.\ (\ref{rhofrommv}) then gives
\begin{equation}
\rho ^{\left( 1/2\right)} = \frac12 \widehat{1} +
\left( w^+_x -\frac12 \right) \widehat{\sigma _x} +
\left( w^+_y -\frac12 \right) \widehat{\sigma _y} +
\left( w^+_z -\frac12 \right) \widehat{\sigma _z},
\label{rhofromw+}
\end{equation}
where we made use of the short notation $w^{\pm}_j \equiv w\left( \pm \frac12,u\left( j\right) \right) $, $j=x,y,z$ and $u\left( j\right) $ the respective $\left( \theta ,\phi \right) $ pair. An analogous expression holds with $w^-_j $. In matrix form, eq. (\ref{rhofromw+}) can be written (with aid of eq.\ (\ref{marginal})):
\begin{equation}
\rho ^{\left( 1/2\right) } =
\left[
\begin{array}{cc}
w^+_z & \left( w^+_x -\frac12 \right) -i\left( w^+_y -\frac12 \right) \\
\left( w^+_x -\frac12 \right) +i\left( w^+_y -\frac12 \right) & w^-_z
\end{array}
\right] .
\label{rhofromwmtx}
\end{equation}

This shows that also $w\left( i,u\right) $ contains redundant information, which are translated as constraints that a function must obey to describe a quantum mechanical state. In this case, the knowledge of three points determine the behaviour of the whole function.

A similar treatment, but which makes use of a nonredundant {\it{quorum}} is made by Weigert\cite{Wei00}. In this treatment, the complete state of a spin $j$ is reconstructed by measuring $4j\left( j+1\right) $ (tomographically) independent spin projection mean values.

For completeness, we reproduce here the general formula which gives density operator from $w\left( i,u\right) $ for any spin value. This formula is deduced in refs.\cite{DM97-MM97-AMMS98} from general properties of Wigner's $3j$  symbols\cite{LL} (equivalently, Clebsch-Gordan coefficients). It reads
\begin{eqnarray}
\left( -1\right)^{m'_2} \sum _{j_3 =0}^{2j} 
\sum _{m_3 =-j_3}^{j_3} \left(2j_3 +1\right) ^2
\sum _{m_1 =-j}^{j} \int \left( -1\right) ^{m_1}
w\left( m_1,\theta ,\phi \right)
D^{\left( j_3\right) }_{0m_3}\left( \phi ,\theta ,\psi \right) \nonumber \\
\times \left(
\begin{array}{ccc}
j&j&j_3 \\
m_1&-m_1&0
\end{array}
\right) \left(
\begin{array}{ccc}
j&j&j_3 \\
m'_1&-m'_2&m_3
\end{array}
\right) d\Omega
=
\rho ^{\left( j\right) }_{m'_1m'_2}, \nonumber
\end{eqnarray}
where $\left( 
\begin{array}{ccc}
j_1&j_2&j_3\\
m_1&m_2&m_3
\end{array}
\right) $
are Wigner's $3j$ symbols, and
\[
\int d\Omega = \frac{1}{8\pi ^2}\int _{0}^{2\pi }d\phi
\int _{0}^{\pi }\sin \theta d\theta \int _{0}^{2\pi } d\psi .
\] 

\subsection{Examples}

In this subsection we compute $w\left(i,u\right) $ for the same four examples given in $P \left( \vec{s} \right) $ representation.

i) Spin up in $z$-axis:
In this case, the rotated density matrix reads
\[
\rho ^{\left( 1/2\right)}\left( u\right) = \left[
\begin{array}{cc}
\cos ^2\frac{\theta}{2} & 
-\sin \frac{\theta}{2}\cos \frac{\theta}{2}e^{i\psi} \\
-\sin \frac{\theta}{2}\cos \frac{\theta}{2}e^{-i\psi} &
\sin ^2\frac{\theta}{2}
\end{array}
\right] ,
\]
and one gets
\begin{eqnarray}
w\left( +\frac12,u\right) = \cos ^2\frac{\theta}{2} = \frac12 \left( 1+\cos \theta \right) ,\nonumber \\
w\left( -\frac12,u\right) = \sin ^2\frac{\theta}{2} = \frac12 \left( 1-\cos \theta \right) . \nonumber
\end{eqnarray}

ii) Spin ``up'' in $x$-axis:
In this case one gets
\begin{eqnarray}
w\left( +\frac12,u\right) = \frac12 \left( 1+\sin \theta \cos \phi \right) ,\nonumber \\
w\left( -\frac12,u\right) = \frac12 \left( 1-\sin \theta \cos \phi \right) .\nonumber
\end{eqnarray}

iii) Spin ``up'' in $y$-axis:
In this case one gets
\begin{eqnarray}
w\left( +\frac12,u\right) = \frac12 \left( 1+\sin \theta \sin \phi \right) ,\nonumber \\
w\left( -\frac12,u\right) = \frac12 \left( 1-\sin \theta \sin \phi \right) .\nonumber
\end{eqnarray}

iv) Unpolarized spin:
Finally, in this case one gets
\begin{equation}
w\left( +\frac12,u\right) = \frac12 = w\left( -\frac12,u\right) .\nonumber
\end{equation}

This last example simply exhibits the isotropy of the state. The other three examples are best understood if one remembers polar spherical coordinates expressions.

In fact, the $w$ distribution for spin $1/2$ can be visualized with aid of the Bloch ball (i.e. Bloch sphere for pure states and its interior for mixed ones). The connection between Bloch vector $\vec{b}$ and density matrix is made by (see eq.\ (\ref{rhofrommv}))
\begin{equation}
\rho ^{\left( 1/2\right)} = \frac12 \widehat{1} + \vec{b} \cdot \vec{\widehat{\sigma }},
\nonumber
\end{equation}
where $\vec{b} \equiv \frac12 \left\langle \vec{\widehat{\sigma}} \right\rangle $ has Euclidean norm lower than or equal to $\frac12$. From the definition of $w$ one obtains
\begin{equation}
w\left( \pm \frac12,u\right) = \frac12 \pm \vec{b} \cdot \vec{u},
\label{wfromb}
\end{equation}
where $\vec{u}$ is the unitary vector defined by the polar angles $\left( \theta ,\phi \right) $.

\section{QUASIPROBABILITY $P$ FROM PROBABILITY $w$}

\label{Pw}In this section we explicit formulas which give the quasiprobability distribution $P\left( \vec{s}\right) $ (in fact the $p\left( c,b,a\right) $ of eq.\ (\ref{p})) from the knowledge of the tomographic probability distribution $w\left( i,u\right) $

The composition of eqs.\ (\ref{pexplict}) and (\ref{rhofromwmtx}) directly give
\begin{eqnarray}
\label{pfromw}
p\left( 1,1,1\right) &=& \frac14 \left( 1+i\right) 
\left[ w^+_x -iw^+_y +w^+_z \right] -\frac14, \\
p\left( -1,1,1\right) &=& \frac14 \left( 1-i\right) 
\left[ -w^+_x +iw^+_y +w^+_z \right] -\frac{i}4, \nonumber \\
p\left( 1,-1,1\right) &=& \frac14 \left( 1-i\right) 
\left[ w^+_x -iw^+_y +w^+_z \right] +\frac{i}4, \nonumber \\
p\left( -1,-1,1\right) &=& \frac14 \left( 1+i\right) 
\left[ -w^+_x +iw^+_y +w^+_z \right] +\frac14, \nonumber \\
p\left( 1,1,-1\right) &=& \frac14 \left( 1-i\right) 
\left[ w^+_x +iw^+_y +w^-_z \right] -\frac14, \nonumber \\
p\left( -1,1,-1\right) &=& \frac14 \left( 1+i\right) 
\left[ -w^+_x -iw^+_y +w^-_z \right] +\frac{i}4, \nonumber \\
p\left( 1,-1,-1\right) &=& \frac14 \left( 1+i\right) 
\left[ w^+_x +iw^+_y +w^-_z \right] -\frac{i}4, \nonumber \\
p\left( -1,-1,-1\right) &=& \frac14 \left( 1-i\right) 
\left[ -w^+_x -iw^+_y +w^-_z \right] +\frac14. \nonumber
\end{eqnarray}  
Those equations are analogous to Radon transform, where Wigner quasiprobability distribution is written down from tomographic probabilities.

In this way, the ``hidden'' quasiprobability $P\left( \vec{s}\right) $ can be obtained directly from the knowledge of tomographic probabilities $w^{\pm}_j$. 

In particular, the four examples worked out previously can be checked, and the knowledge of the three Cartesian axis tomographic probabilities allows one to completely determine the ``joint quasiprobability'' $P$.

\section*{Concluding Remarks}

We have developed a Radon like transformation, in which $P$ quasiprobability distribution for spin $1/2$ states is written in terms of (some special points of) the tomographic probability distribution $w$.

The theme of (quasi)probability distributions for spin states is much more rich, and besides these two treated here, there are some others ``Wigner'' distributions proposed by diverse authors, with different motivations. As an example, in ref.\cite{CKW00}, Chumakov, Klimov and Wolf show that a Wigner quasiprobability distribution proposed by Agarwal\cite{Aga81} is a restriction of a Wigner function later proposed by Wolf\cite{Wol96} for generic Lie groups.

Some peculiar properties of $P$ quasiprobability are evidenced. They are related to the operator ordering question, also addressed in ref.\cite{Cha92}.

The connection between Bloch vector and $w$ tomographic distribution is also made (eq.\ (\ref{wfromb})). It shows a direct geometric interpretation of $w$ in the case of two level systems.

\section*{Acknowledgement}

Authors thank organizers of XXXI Latin American School of Physics (ELAF '98), held in Mexico city, in honor of Professor Marcos Moshinsky, where part of this work was done, for hospitality, and one anonymous referee for suggestions.

\section*{Appendix: Calculational tools}

In this work we made use of Pauli spin matrices in the form 
\[
\widehat{\sigma _x}=\left( 
\begin{array}{cc}
0 & 1 \\ 
1 & 0
\end{array}
\right) ,
\widehat{\sigma _y}=\left( 
\begin{array}{cc}
0 & -i \\ 
i & 0
\end{array}
\right) , 
\widehat{\sigma _z}=\left( 
\begin{array}{cc}
1 & 0 \\ 
0 & -1
\end{array}
\right) . 
\]
Conventionally,
\begin{eqnarray}
\left| \pm x\right\rangle &=&
\frac 1{\sqrt{2}} \left\{
\left| z\right\rangle \pm \left| -z\right\rangle
\right\} ,\nonumber \\
\left| \pm y\right\rangle &=&
\frac 1{\sqrt{2}} \left\{
\left| z\right\rangle \pm i\left| -z\right\rangle
\right\} ,\nonumber 
\end{eqnarray}
and follows
\[ 
\begin{array}{ccccc}
\left\langle +x\mid +y\right\rangle \left\langle +y\mid +z\right\rangle
\left\langle +z\mid +x\right\rangle &=&\left\langle -x\mid -y\right\rangle
\left\langle -y\mid +z\right\rangle \left\langle +z\mid -x\right\rangle &=& \\
\left\langle -x\mid +y\right\rangle \left\langle +y\mid -z\right\rangle
\left\langle -z\mid -x\right\rangle &=&\left\langle +x\mid -y\right\rangle
\left\langle -y\mid -z\right\rangle \left\langle -z\mid +x\right\rangle
&=&\frac 14\left( 1+i\right) , \\
\left\langle -x\mid -y\right\rangle \left\langle -y\mid -z\right\rangle
\left\langle -z\mid -x\right\rangle &=&\left\langle +x\mid +y\right\rangle
\left\langle +y\mid -z\right\rangle \left\langle -z\mid +x\right\rangle &=& \\
\left\langle +x\mid -y\right\rangle \left\langle -y\mid +z\right\rangle
\left\langle +z\mid +x\right\rangle &=&\left\langle -x\mid +y\right\rangle
\left\langle +y\mid +z\right\rangle \left\langle +z\mid -x\right\rangle
&=&\frac 14\left( 1-i\right) , \\
\left\langle +x\mid +y\right\rangle \left\langle +y\mid +z\right\rangle
\left\langle -z\mid +x\right\rangle &=&-\left\langle -x\mid -y\right\rangle
\left\langle -y\mid +z\right\rangle \left\langle -z\mid -x\right\rangle &=&  \\
-\left\langle -x\mid +y\right\rangle \left\langle +y\mid -z\right\rangle
\left\langle +z\mid -x\right\rangle &=&\left\langle +x\mid -y\right\rangle
\left\langle -y\mid -z\right\rangle \left\langle +z\mid +x\right\rangle
&=&\frac 14\left( 1+i\right) , \\
\left\langle +x\mid -y\right\rangle \left\langle -y\mid +z\right\rangle
\left\langle -z\mid +x\right\rangle &=&-\left\langle -x\mid +y\right\rangle
\left\langle +y\mid +z\right\rangle \left\langle -z\mid -x\right\rangle &=& \\
-\left\langle -x\mid -y\right\rangle \left\langle -y\mid -z\right\rangle
\left\langle +z\mid -x\right\rangle &=&\left\langle +x\mid +y\right\rangle
\left\langle +y\mid -z\right\rangle \left\langle +z\mid +x\right\rangle
&=&\frac 14\left( 1-i\right) . 
\end{array}
\]

\end{document}